# Influence of Atomic Roughness at The Uncompensated Fe/CoO (111) Interface on Exchange Bias Effect

*Rui Wu*[1,3,4†], *Mingzhu Xue*[1,3†], *Tuhin Maity*[4], *Yuxuan Peng*[1,3], *Samir Kumar Giri*[4], *Guang Tian*[1,3], *Judith L. MacManus-Driscoll*[4]* *and Jinbo Yang*[1,2,3]*

1．State Key Laboratory for Mesoscopic Physics, School of Physics, Peking University, Beijing 100871, P. R. China

2．Collaborative Innovation Center of Quantum Matter, Beijing 100871, P. R. China

3．Beijing Key Laboratory for Magnetoelectric Materials and Devices，Beijing 100871, P. R. China

4．Department of Materials Science and Metallurgy, University of Cambridge, Cambridge, CB3 0FS, United Kingdom

†These authors contribute to this work equally.

* Corresponding author: jld35@cam.ac.uk and jbyang@pku.edu.cn.

**Abstract**

The effect of interface roughness of ferromagnetic and antiferromagnetic layers on exchange bias is still not well understood. In this report we have investigated the effect of surface roughness in (111)-oriented antiferromagnetic CoO films on exchange bias with ferromagnetic Fe grown on top. The surface roughness is controlled at the atomic scale, over a range below ~ 0.35 *nm*, by varying layer thickness of the CoO films. It is observed that both exchange bias field ($H_E$) and coercivity ($H_C$) extensively depend on the atomic scale roughness of the CoO (111) at the interface with Fe film. An opposite dependence of $H_E$ and $H_C$ on interface roughness was found, which was ascribed to partially compensated spin states induced by the atomic roughness at the fully uncompensated CoO (111) surfaces and was corroborated using the Monte Carlo simulations. Moreover, the onset temperature for $H_C$ is found to be up to ~ 80 *K* below the blocking temperature ($T_B$) and the temperature dependence of $H_C$ follows the power law with a critical exponent equal to one, which indicates that, in this system, $H_C$ is more of an interface-related property than $H_E$.

## I. INTRODUCTION

The exchange bias (EB) effect in ferromagnetic (FM)/antiferromagnetic (AFM) heterogenous structures is one of the most studied magnetic phenomena, whereby a shift of the hysteresis loop along the field axis is observed either in the positive or negative direction after field cooling (FC) through the Néel temperature ($T_N$) of the AFM layer [1-6]. It has been studied for decades due to the wide spectrum of applications which rely on it, such as giant magneto resistance (GMR) read heads and various magnetic random-access memories (MRAM). In addition to the loop shift, several other interesting properties such as the enhancement of coercivity ($H_C$) for nanoscale FM materials are also observed which can overcome the superparamagnetic limit in future ultra-high-density recording media [7] and improve the energy product of hard magnetic materials [8].

As an interface effect, both EB field ($H_E$) and $H_C$ show significant dependence on interface roughness [2]. Despite intensive research on this, most of the investigations are focused on the polycrystalline AFM films where the crystalline orientation was ignored [9, 10] while the effect of roughness significantly influences the interface spin compensation for EB coupling [11]. Even different origins of roughness for the same materials can significantly influence the FM/AFM interface and give different EB. For instance, in cleaved uncompensated CoO (111) surfaces, both $H_E$ and $H_C$ can be increased by interfacial roughness induced by polishing [12]. However, in Co films with natively oxidized (111)-textured CoO layers on top, a decreased $H_E$ and an increased $H_C$ are obtained with increasing interface roughness induced by longer oxidizing times [13]. The common methods to create interfacial roughness such as polishing [12], oxidizing films [13], ionic bombardment of the surface [14], and higher growth temperature [15] are destructive and can inevitably change the bulk properties of AFM layer at the interface and the stoichiometry of the materials. Moreover, in previous studies, an interface roughness of up to several nanometers was induced in the AFM layer [14,15]. The magnetic properties of the FM layer would also have been changed with such a rough interface, thus introducing another variable. Hence, a non-destructive technique to control the interface roughness at an atomic level is necessary to clearly elucidate the influence of interface roughness on EB.

In this work, the roughness of highly textured CoO (111) films was carefully controlled from ~0.25 *nm* to ~0.35 *nm* using pulsed laser deposition (PLD). This was done by using CoO layers of different thickness ($t_{CoO}$), i.e. 5 *nm*, 10 *nm*, 15 *nm*, 20 *nm*, 30 *nm*, and 40 *nm*, in the Fe/CoO(111) bilayers. With increasing $t_{CoO}$, an increasing surface roughness of CoO film was obtained. Compared to the aforementioned methods, film thickness control of roughness is not destructive. Hence, the chemical stoichiometry and thus the related magnetic properties of the AFM layer are not expected to change significantly with increasing thickness. A drastic increase of $H_C$ and decrease of $H_E$ with increasing CoO interface roughness was observed over a wide temperature range. Monte Carlo simulations were also undertaken, and they were in good agreement with the experimental results, proving the roughness-controlled interface spin compensation state is linked to the EB. Moreover, the temperature dependency of $H_C$ and $H_E$ show different origins to these two quantities, i.e. $H_C$ is more interface-related whereas $H_E$ depends on both bulk the spins and the interfacial spins of the AFM layer.

## II. EXPERIMENTAL PROCEDURE

Bilayer samples of Fe/CoO (111) were fabricated using pulsed laser deposition (PLD, λ = 248 *nm*) with a pulse repetition rate of 5 *Hz*, and a spot energy density of 2 *J/cm²* at room temperature. The highly-textured CoO (111) films of thickness 5 *nm*, 10 *nm*, 15 *nm*, 20 *nm*, 30 *nm*, 40 *nm*, and 50 *nm* were reactively deposited from a cobalt target (99.99 %) onto a Si/SiO$_2$ substrates under the oxygen pressure of 2.3×10$^{-4}$ *Torr*. Then, an Fe layer of 5 *nm* was deposited under an Ar pressure of 1.8×10$^{-3}$ *Torr* and capped with 3 *nm* Au to prevent oxidation. The microstructure of the CoO (50 *nm*)/Fe (5 *nm*)/Au (3 *nm*) sample was investigated using high-resolution transmission electron microscopy (HRTEM). Magnetic measurements were carried out for all samples with a CoO thickness from 5 *nm* to 40 *nm* using the longitudinal magneto-optical Kerr effect (L-MOKE). To study the crystalline structure and surface morphology using X-ray diffraction (XRD) and atomic force microscopy, another batch of the bare CoO films of thickness 10 *nm*, 20 *nm*, 40 *nm*, and 80 *nm* were prepared.

## III. RESULTS AND DISCUSSION

The evolution of the roughness of the bare CoO surface with increasing film thickness was demonstrated using atomic force microscopy. As shown in Fig. 1(A), the surface is smooth for the 10 *nm* film while the roughness is increased for samples with higher thickness (Fig. 1(B-D)). The highest roughness is obtained in the surface of the 80 *nm* CoO sample (Fig. 1(D)). As summarized in Fig. 1(E), the RMS roughness ($\sigma$) calculated from the atomic force microscopy images. The lateral correlation length ($\xi$), which is a measure of the size of the mounds for a mounded surface, was estimated by fitting an exponential to the autocorrelation function of the profile data [16, 17]. It was found that $\sigma$ increases steadily from 2.6 Å to 3.7 Å, while $\xi$ increases from 2.4 nm to 10.5 nm, with film thickness increase from 10 *nm* to 80 *nm*. Among these two parameters, the RMS roughness is related to the magnitude of the surface fluctuation and thus determines how many atomic planes in the antiferromagnetic layer can be exchange coupled to the top ferromagnetic layer. In Fig. 1(F), the XRD results of bare CoO films with two typical thicknesses of 20 *nm* and 40 *nm* are shown. A clear peak of CoO (111) is observed in both films, indicating a highly (111)-oriented texture in the CoO layer. It is known that bulk CoO has a type-II AFM structure, where the magnetic moments of the Co atoms in the same (111) plane are aligned in parallel, while the adjacent planes are coupled antiferromagnetically [18]. Much larger EB values have been demonstrated in CoO (111)/FM bilayers compared to CoO (001)/FM bilayers owing to an uncompensated *vs* compensated interface [*19*]. In the XRD data, no obvious change of the peak position and peak width was observed apart from obvious increase of intensity with film thickness, which confirms that the crystalline quality does not change significantly with increasing film thickness in our films. Moreover, a tiny peak located at ~44.2° in the 40 nm CoO film, likely indicates the existence of a small amount of (001)-oriented CoO.

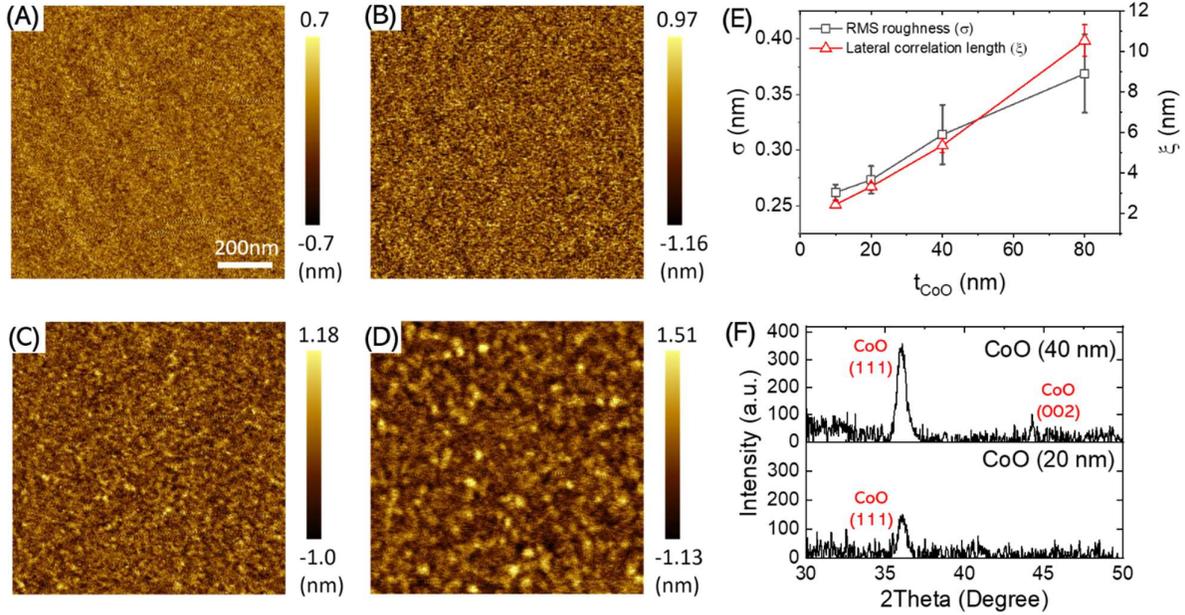

**FIG. 1. (A-D) Atomic force microscopy images for CoO films of 10 *nm*, 20 *nm*, 40 *nm*, and 80 *nm*, respectively. (E) The dependence of the RMS roughness and the lateral correlation length averaged from 4 different area of the CoO film surfaces on the film thicknesses. The error bars are the standard deviations from the average. (F) XRD patterns of CoO films with two typical thicknesses of 20 *nm* and 40 *nm*.**

Cross-sectional HRTEM characterization was carried out for a CoO (50 *nm*)/Fe film. As shown in Fig. 2 (A), the CoO, Fe, and Au layers can be easily distinguished with relatively sharp interfaces. The CoO layer shows a complex structure, where some CoO grains grow larger to form a columnar structure while some grains shrink during the growth and form a "△" grain. The lattice fringes in the columnar grains give a lattice distance of ~2.55 *nm*, confirming the (111) orientation of the CoO film. The "△" grain structure is more clearly seen in Fig. 2(B). A weaker lattice fringes in this area might be related to the either poorer crystalline in a smaller grain (< 5 *nm*) or larger strain at the grain boundary. Such a microstructure is usually observed in high melting point materials deposited at temperatures below ~ 0.2-0.3 $T_m$, where $T_m$ denotes the melting temperature [20]. Here, our growth temperature ($T_g$ ~ 300 K) is also much lower than the melting temperature of CoO ($T_m$ ~ 2206 K, $T_g/T_m$ ~ 0.14). It should be noted that the lateral grain size is large compared to the scale of atomic surface roughness. Hence, we can reasonably assume that the change of lateral size of the CoO grain with film thickness will not significantly affect the EB in our system. Moreover, the interface roughness in the TEM image shown in Fig. 2(A) agrees well with that obtained from surface characterization by atomic force microscopy. It is observed that the surface fluctuation of the layer can cross the grain boundaries, indicating that correlation length and grain size cannot be directly linked to each other.

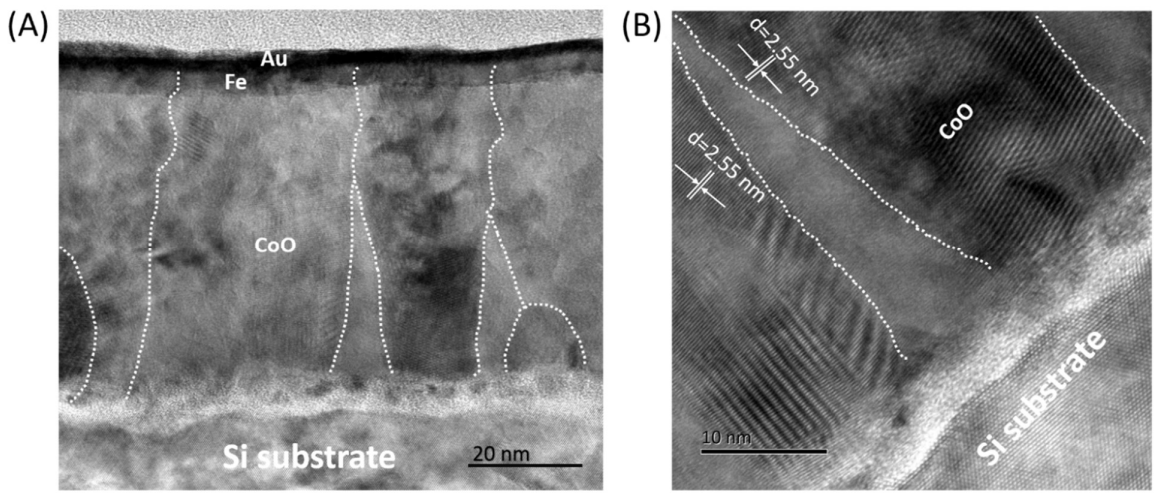

**FIG. 2. (A) The cross-sectional TEM and (B) HRTEM images of a Fe/CoO (111) film with $t_{CoO}$ of ~50 *nm*. The dotted lines indicate the grain boundaries in the CoO layer.**

For the EB measurements, the samples were first field-cooled (FC) from 330 *K* to 80 *K* in an external field of 500 *Oe*, which is much higher than the saturation field (only several tens of Oersted) of the samples. Then hysteresis loops were measured with temperature varied from 80 *K* to 330 *K*. Two consecutive hysteresis loops were measured at 80 *K* for each sample after field cooling, and no training effect was observed in all these samples. Thus, consecutive hysteresis loop measurements with varying temperature were carried out with FC only once. The temperature dependent behaviors are shown in Fig. 3(A) and Fig. 3(B) for CoO (5 *nm*)/Fe and CoO (40 *nm*)/Fe, respectively. The CoO (5 *nm*)/Fe film shows a blocking temperature ($T_B$, the temperature where the loop shift becomes zero) of 280 *K* and the whole hysteresis loops shift to the negative field direction with the decrease of temperature, i.e., both the left coercive field ($H_{CL}$) and right coercive field ($H_{CR}$) are negative, at all temperatures below $T_B$. For the CoO (40 *nm*)/Fe film, a $T_B$ of 325 *K* is observed, which is significantly higher than the Néel temperature of bulk CoO ($T_N$ = 292 K). The enhancement of the $T_N$ is related to the presence of oxygen vacancies in the CoO film, which modifies the superexchange interaction in the CoO film [*21*]. At temperatures lower than 240 *K*, in contrast to the CoO (5 *nm*)/Fe film, the hysteresis of the CoO (40 *nm*)/Fe film partially shifts to the negative field direction, showing negative $H_{CL}$ and positive $H_{CR}$, only at temperatures close to $T_B$ does the whole hysteresis loops shift in the negative field quadrant with both $H_{CL}$ and $H_{CR}$ negative. The hysteresis loops of samples with different $t_{CoO}$ measured at 80 *K* are shown in Fig. 3(C). A unidirectional enhancement of $H_C$ is observed with increasing $t_{CoO}$, i.e., a drastic increase is only found in the right branches of the hysteresis loops while the left branches remain nearly unchanged.

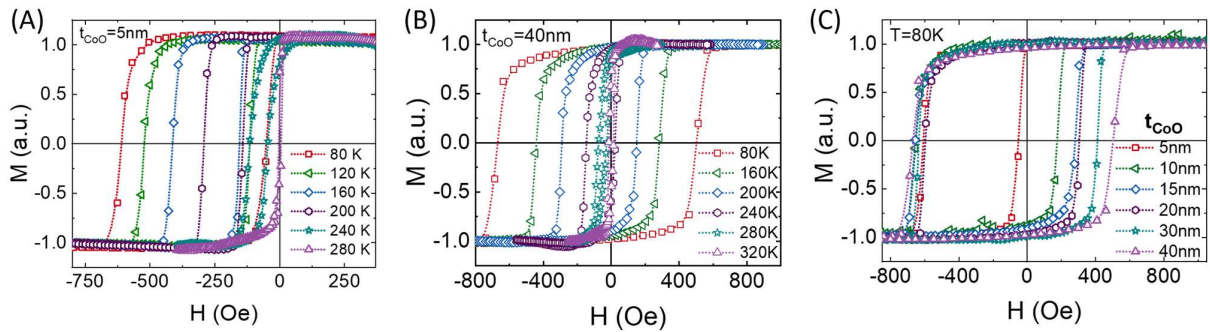

**FIG. 3.** Hysteresis loops of (A) CoO (5 *nm*)/Fe and (B) CoO (40 *nm*)/Fe films measured at different temperatures and (C) hysteresis loops at 80 *K* of Fe/CoO($t_{CoO}$) with different $t_{CoO}$. All the hysteresis loops were measured after FC with an external field of 500 *Oe*.

The temperature evolution of $H_{CL}$ and $H_{CR}$ of the hysteresis loops are given in Fig. 4(A) for all the samples. It is found that $H_{CL}$ of all the samples are close to each other, showing a linear dependence with temperature and nearly no dependence with $t_{CoO}$. However, the temperature dependence of $H_{CR}$ is significantly different for each sample. As temperature decreases, $H_{CR}$ first changes towards the negative field direction, following $H_{CL}$, and then turns to the positive direction, giving rise to a bifurcation in both curves.

Fig. 4(B) shows the temperature dependence of $H_C = (H_{CR}-H_{CL})/2$, and the EB field, $H_E = (H_{CR}+H_{CL})/2$, of samples with different $t_{CoO}$. For all the samples, $H_E$ tends to saturate at low temperatures. However, the thicker films tend to saturate more quickly compared to thinner films due to the quick stabilization of the AFM moment below $T_B$, in good agreement with theory proposed by Stiles *et al.* [*22*]. Except for $H_E$, the exchange coupling with the CoO layer also gives an enhancement to the $H_C$ of the Fe film at low temperatures. The onset of the $H_C$ enhancement and the onset of the $H_E$ occur at different temperatures. Here, we define $T'_B$ as the temperature where the enhancement of the $H_C$ happens and we call it the blocking temperature for $H_C$, analogous to the blocking temperature for $H_E$ (denoted with $T_B$). It is noticed that, in the temperature below $T'_B$, the $H_C$ shows a linear dependence of the temperature for all these samples. Thus, we can determine the $T'_B$ with a linear fitting $H_C \sim 1-T/T'_B$ ($T < T'_B$).

To prove that the unique $t_{CoO}$ dependence observed here does not originate from the evolution of microstructure of CoO non-epitaxially grown on Si/SiO$_2$ substrates, the controlled experiments with growing Fe/CoO(111) films on MgO(111) substrates under the same condition were carried out. Since CoO and MgO have similar lattice parameters, i.e. $a_{CoO}$=0.426 nm and $a_{MgO}$=0.421 nm, the growth is easily epitaxial even at room temperature [23]. The supplementary Figure S1(A) and S1(B) shows the exchange biased magnetic hysteresis loops measured with the same procedure with Fe/CoO(111) samples grown on Si/SiO$_2$ substrates. The magnetic field was applied along the [01-1] direction during the FC and hysteresis loop measurement. It is found that, with the $t_{CoO}$ increasing from 5 nm to 30 nm, the $H_E$ decreases from -274 Oe to -120 Oe while the $H_C$ increases from 412 Oe to 598 Oe at 80 K, as shown in supplementary Figure S2. Since the epitaxial growth of CoO(111) on MgO(111) substrate can be expected, which will exclude the existence of pyramid grain structure as well as the (001)-orientated grains in CoO films, the consistence between the magnetic results with those of samples grown on Si/SiO$_2$ substrates demonstrates that the microstructure of the bulk CoO is not responsible for the magnetic properties observed here. Moreover, as shown in Fig. S3, a similar thickness dependence of the surface RMS roughness was observed in the CoO films grown on MgO(111) substrates. This indicates the film thickness dependence of the surface morphology does not depend on the substrates used in the deposition.

Fig. 4(C) compares $H_C$ and $H_E$ of films with different $t_{CoO}$ measured at 80 K. As $t_{CoO}$ increases, $H_E$ shows a gradual decrease from -350 Oe in the sample with 5 nm CoO to only -90 Oe in the sample with 40 nm CoO, corresponding to a change of the interface exchange energy density ($\sigma_{ex} = M_{FM} t_{FM} H_E$) from 0.281 erg/cm$^2$ to 0.071 erg/cm$^2$. Meanwhile, with increasing $t_{CoO}$, $H_C$ shows a gradual increase from 280 Oe to 587 Oe in films with $t_{CoO}$ = 5 nm and 40 nm, respectively. Both $H_E$ and $H_C$ can be fit with exponential functions, $H_{E(C)}(t_{CoO}) = A\exp(-t_{CoO}/t_0) + H_{E(C)}^{\infty}$, yielding $t_0$ = 12.7±6.8 nm, $H_C^{\infty}$ = 580±57 Oe for $H_C$ and $t_0$ = 12.1±0.9 nm, $H_E^{\infty}$ = -73.4±7.4 Oe for $H_E$. It is found that this *opposing* dependence of $H_E$ and $H_C$ on $t_{CoO}$ is not significant in the metallic FM/AFM EB systems, such as IrMn(111)/FM and FeMn(111)/FM systems, where $H_E$ and $H_C$ nearly do not change with AFM layer thickness larger than 10 $nm$ [24,25], although an AFM-thickness-dependent interfacial roughness was also observed in these systems [26]. However, it should be noted that a noncollinear 3Q antiferromagnetic structure is present in these systems. Thus, the pristine IrMn(111) and FeMn(111) surfaces are not perfectly uncompensated, which is different from the CoO(111) surface discussed here.

As shown in Fig. 4(D), a large difference between $T_B$ and $T'_B$ is observed, which is up to 80 $K$ for the sample of thickness $t_{CoO} = 10$ $nm$. Moreover, a distinct dependence on $t_{CoO}$ is also observed for $T_B$ and $T'_B$. $T_B$ initially increases with increasing $t_{CoO}$ and then saturates at $t_{CoO} = 10$ $nm$ and thereafter. This agrees well with the thickness dependence of $T_N$ in CoO thin films, where such a saturation was also obtained in films thicker than 10 $nm$ [27]. Since the $T_B$ is very sensitive to the stoichiometry of the CoO [21], this indicates that the stoichiometry of the CoO layers keeps unchanged with increasing $t_{CoO}$. On the other hand, $T'_B$ keeps increasing with an increase of $t_{CoO}$. The $t_{CoO}$ dependency of both $T_B$ and $T'_B$ can be fit very well with the finite-size scaling formula in the ultrathin limit, which have been widely used to describe the AFM thickness dependence of the blocking temperature and give a quantitative description of phase transition from ordered state to disorder state [28,29]:

$$\frac{T_{B,\infty} - T_B(t_{AFM})}{T_{B,\infty}} = \left(\frac{\xi_0}{t_{AFM}}\right)^\delta, \quad (1)$$

$$\frac{T'_{B,\infty} - T'_B(t_{AFM})}{T'_{B,\infty}} = \left(\frac{\xi'_0}{t_{AFM}}\right)^{\delta'}, \quad (2)$$

where $\xi_0 = J_{INT}^{(0)} / 2K_{AFM}^{(0)} D a_0$ is the coherent length with $J_{INT}^{(0)}$, $K_{AFM}^{(0)}$, $D$, and $a_0$ being the interfacial exchange coupling and anisotropy energy of AFM spins at T = 0, the lateral dimension of AFM grain, and the distance between two AFM spins, respectively. $\xi_0$ denotes the thickness of the AFM interface layer contributing to $H_E$. $T_{B,\infty}$ is the blocking temperature for infinite AFM layer thickness ($t_{AFM} \to \infty$) and $\delta$ is the critical exponent. The parameters $T'_{B,\infty}$, $\xi'_0$ and $\delta'$ are defined accordingly for $H_C$ and $T'_B$.

From the fitting shown in Fig. 4(D), the exponential indexes are obtained which are $\delta = 3.17$ and $\delta' = 0.55$ for $T_B$ and $T'_B$, respectively. $T_{B,\infty} = 325.4\ K$ and $T'_{B,\infty} = 290.2\ K$ are also obtained, indicating a difference of 35 $K$ still exists between these two characteristic temperatures at $t_{AF} \to \infty$. Previous work has shown that $T_B$ should converge to $T_N$ of the bulk AFM film, i.e. $T_{B,\infty} = T_N$ [30]. Thus, a much lower $T'_{B,\infty}$ than $T_N$ indicates different origins for $H_E$ and $H_C$ in our system. Meanwhile, the characteristic lengths obtained are $\xi_0 = 2.7\ nm$ and $\xi'_0 = 0.35\ nm$. A smaller $\xi'_0$ than $\xi_0$ is obtained mainly due to smaller $J^{(0)}_{INT}$ for $H_C$ than that for $H_E$. This is because the AFM spins, responsible for $H_C$ are largely compensated, leading to reduction of $J^{(0)}_{INT}$. This result indicates that the length scale of AFM spins underneath the interface that contribute to $H_C$ is much smaller than that of AFM spins contributing to $H_E$. Hence, $H_C$ is more of an interface related property than $H_E$ is. This agrees with the conjecture that $H_C$ is influenced by either interface spin frustration [31] or spin-flop coupling between the FM spins to the compensated interface spins induced by atomic roughness [32].

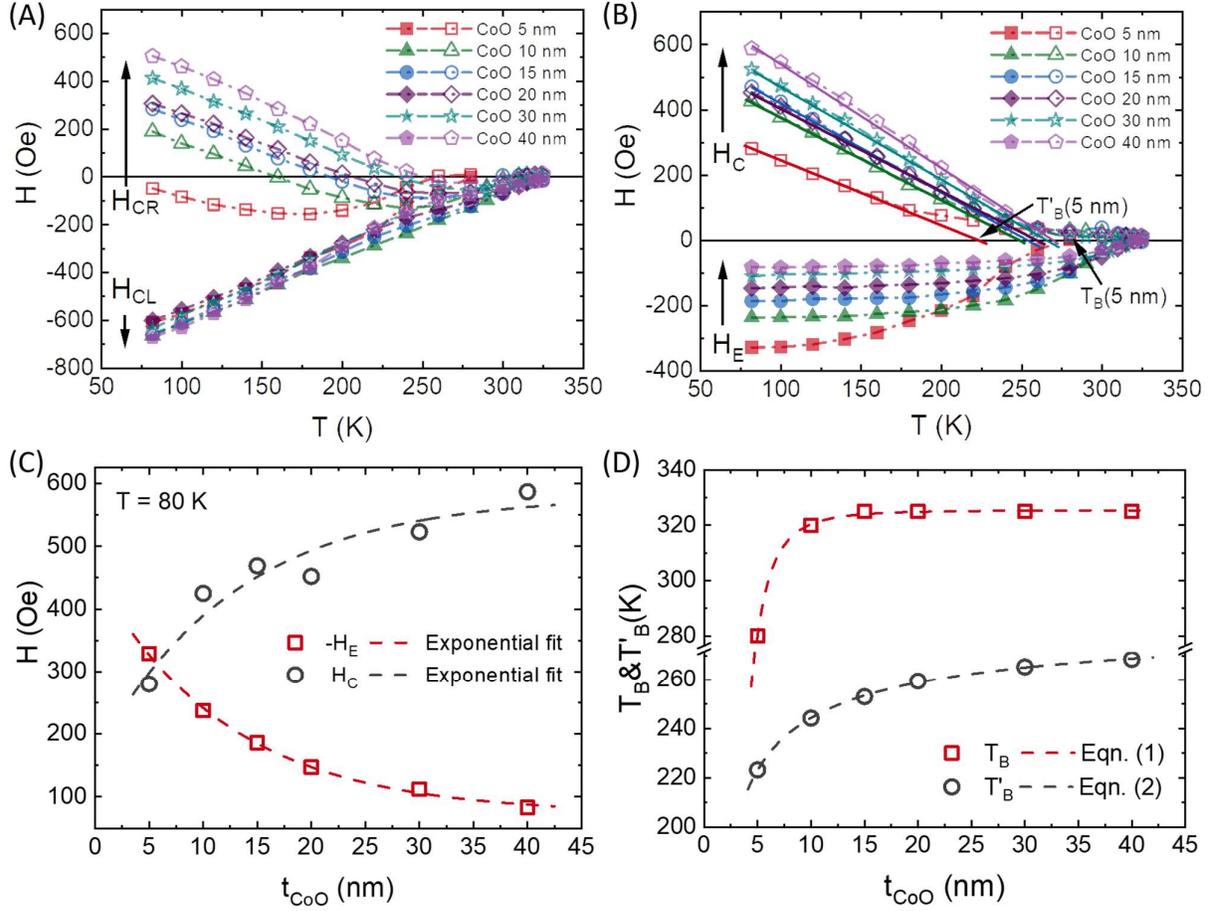

FIG. 4. (A) The temperature dependence of $H_{CL}$ and $H_{CR}$, and (B) the temperature dependence of $H_C$ and $H_E$ in CoO (111)/Fe thin films with different $t_{CoO}$. (C) The dependence of $H_E$ (80 K) and $H_C$ (80 K) on $t_{CoO}$. The dashed lines are the exponential fits. (D) The dependence of $T_B$ and $T'_B$ on $t_{CoO}$. The dashed lines are fit using Eqn. (1) and Eqn. (2).

The interface characteristic of $H_C$ is elucidated by determination of surface critical exponent $\beta_s$, which is defined by the power law [33],

$$\langle M_S \rangle \sim t^{\beta_s}, (3)$$

where $<M_S>$ denotes the order parameter of AFM interface spins responsible for the enhanced $H_C$, and $t = 1 - T/T_C$ is the reduced temperature. Here, $T_C$ can be replaced by $T'_B$ of $H_C$. The evolution of the $H_C$ with temperature reflects the evolution of the order parameter $<M_S>$, which should disappear at $T'_B$. Thus, $H_C$ is supposed to be proportional to $<M_S>$, i.e. $H_C \sim t^{\beta_s}$.

Fig. 5(A) shows a log-log plot of $H_C \sim t^{\beta_s}$ for all samples. The data are fitted well using a power law. The values of $\beta_s$ is given in Fig. 5(B) as a function of $t_{CoO}$. $\beta_s$ shows a linear increase from 0.95 in the sample with 5 *nm* CoO to 1.05 in sample with 40 *nm* CoO which is very close to the value obtained from the mean-field theory for a Heisenberg antiferromagnet, where $\beta = 0.5$, $\beta_s = 1$ for the bulk spins and the surface spins, respectively [34]. Smaller values for $\beta_s$ are usually obtained in experimental studies probably due to the fact that the contribution from interface spins cannot be separated very well with that from the bulk spins. In FeF$_2$-Fe bilayers, $\beta_s \sim 0.8 \pm 0.04$ was obtained from $H_E \sim t^{\beta_s}$ [33], and in the NiO (001) surface, $\beta_s \sim 0.89 \pm 0.01$ was obtained with exchange reflection in low energy electron diffraction (LEED) [35]. The values of $\beta_s \sim 1 \pm 0.05$ obtained from $H_C \sim t^{\beta_s}$ in our system are very close to the theoretical value and larger than those previously obtain from $H_E \sim t^{\beta_s}$, indicating that $H_C$ in this system shows stronger correlation to the roughness induced magnetic frustration or spin-flop coupling from compensated interface spins, than to the bulk spins. This is different to $H_E$, which is typically dominated by both interface and bulk spins [36]. Moreover, the relation $\beta_s \sim 2\delta'$ is in good agreement with theory [28]. The decrease of $\beta_s$ with decreasing $t_{CoO}$ shown in Fig. 5(B) agrees well with that the bulk spin contribute to $H_C$ more for lower AFM film thickness *via* superparamagnetic fluctuation [22].

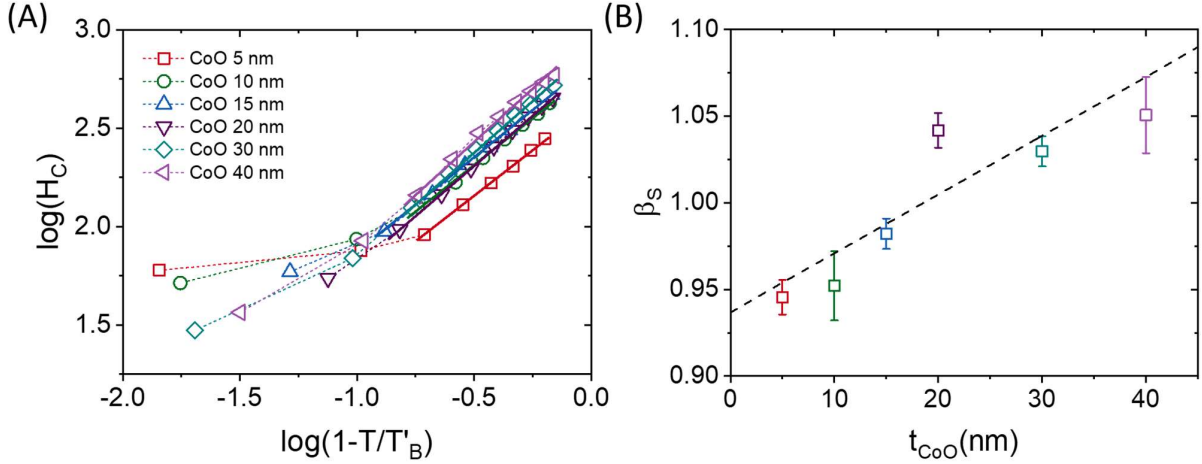

FIG. 5. (A) Log-log graph of $H_C$ vs $t$. The straight line is the plot of the power law $H_C \sim t^{\beta_S}$. (B) $\beta_S$ obtained from (A), as a function of $t_{CoO}$. The straight line is the linear fit.

## IV. MONTE CARLO SIMULATIONS

To further understand the anomalous behaviors of $H_E$ and $H_C$ in our system, atomic Monte Carlo simulations of the EB were carried out for the FM/AFM bilayers which contain uncompensated interfaces with different roughness. A model system with FM thickness, AFM thickness and lateral size of $t_{FM} = 10a$, $t_{AFM} = 20a$, and $l = 50a$, respectively, was considered where $a$ is the unit cell size of a simple cubic lattice. Atoms in the AFM layer that are directly exchanged coupled to the FM layer are defined as the interface. Taking $a = 0.3$ *nm*, such a system corresponds to typical real dimensions $t_{FM} = 3$ *nm*, $t_{AFM} = 6$ *nm*, and $l = 15$ *nm,* with 75000 spins in total. The interface roughness is added by intermixing atoms between FM bottom layer and AFM top layer randomly with a probability $P_{mix}$ varying from 0 to 0.3. It is assumed that the FM atoms cannot go into the bulk AFM. Thus, our algorithm for the atom intermixing has avoided the formation of the isolated FM or AFM atoms at the interface. This can be achieved with two steps. We assume $L_1$ and $L_2$ are the FM and AFM layers at the interface. In the first step, we change atoms in $L_1$ to AFM atoms (then included in $L_2$) with probability $P_{mix}$. We record the position set (*i, j*) of those atoms, with $i$ and $j$ denoting the in-plane atom coordinates. In the next step, in order to keep the total number of atoms for both FM and AFM layers unchanged after atoms exchange, we need to change the same number of atoms in $L_2$ to FM atoms (then included in $L_1$). However, those AFM atoms cannot be selected in the previous position set ($i, j$), which is the key point for avoiding the isolated atoms. If the AFM atoms are selected inside the set ($i, j$), isolated atoms will be created. Fig. 6 (A) and (B) shows three typical bilayers with different $P_{mix}$ values. It is clearly seen that with increasing $P_{mix}$, the roughness of the interface is significantly increased, penetrating across two atomic layers. Fig. 6 (C) shows a schematic spin configuration at the rough interface after field cooling. In Fig. 6(D), the calculated RMS roughness σ of interface with different $P_{mix}$ is given. It indicates that a monotonic increase of interface roughness can be induced by the increase of $P_{mix}$.

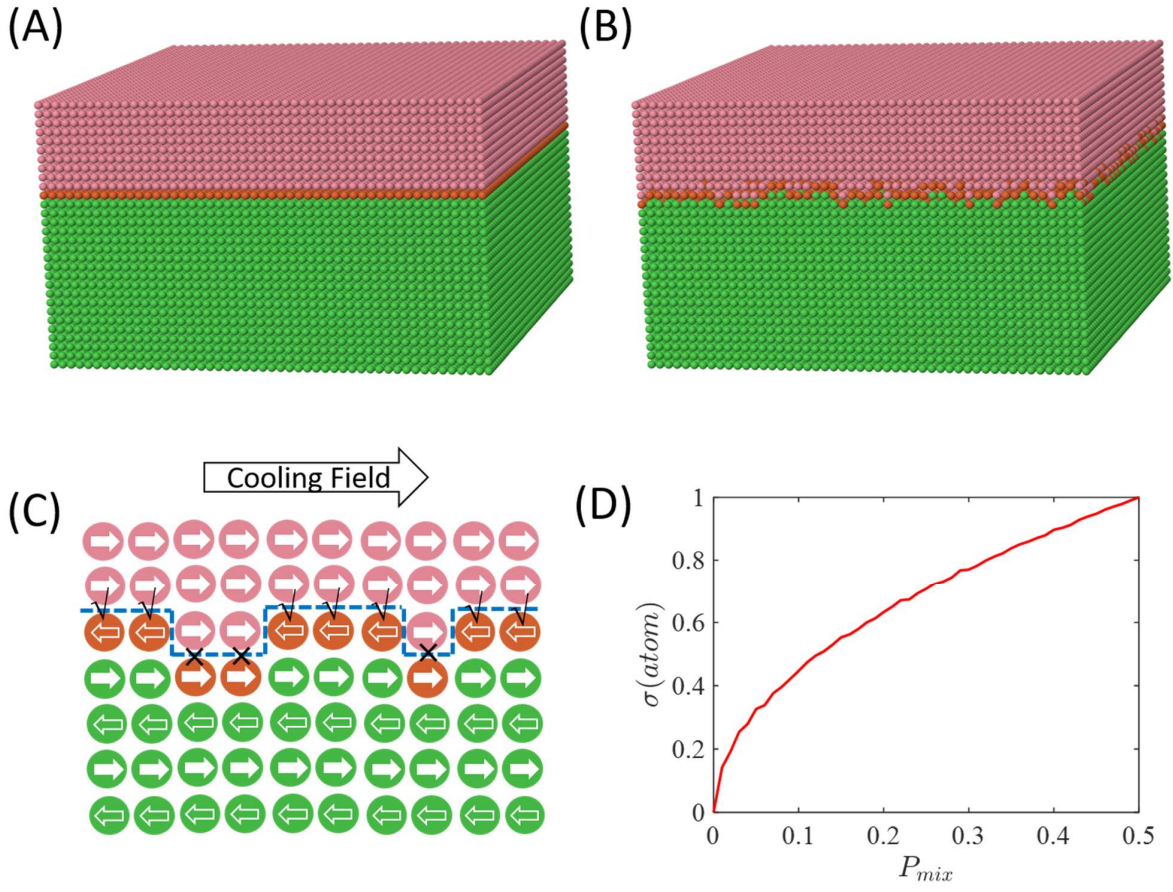

**FIG. 6. FM/AFM bilayers with two typical intermixing probabilities of (A) $P_{mix}$ = 0 and (B) $P_{mix}$ = 0.3. (C) Schematic of the magnetic structure in the FM/AFM bilayer with a rough**

**uncompensated interface. The pink, green, and orange atoms compose the FM layer, AFM bulk, and AFM surface (i.e. interface) respectively. The " √ " and "✕" symbols denote the energy of favorable and unfavorable alignments respectively of the spins with a negative interfacial exchange coupling. (D) The calculated RMS roughness of the interface with different atomic intermixing.**

The anisotropic Heisenberg spin model is adopted in the calculations with a Hamiltonian given by $\mathcal{H} = -\sum_{\langle i,j \rangle} J_{ij} \vec{S}_i \cdot \vec{S}_j - \sum_i K_i (\vec{S}_i \cdot \vec{e}_i)^2 - \sum_i \vec{H} \cdot \vec{S}_i$ where $\vec{S}_i$ are classical Heisenberg spins of unit magnitude placed at the nodes of the simple cubic lattice. The first term represents the exchange energy between spins located in the FM layer, AFM layer and the FM/AFM interface with exchange coupling constants $J_{ij}$ being $J_{FM}$, $J_{AFM}$ and $J_{INT}$, respectively. Here a positive intralayer $J_{AFM}$ is used to yield a parallel alignment of spins in the uncompensated crystalline planes while a negative interlayer $J_{AFM}$ enables the antiparallel alignment of spins in different magnetic sublattices [37]. The second term gives the local anisotropic energy for each spin in the FM and AFM layers with the anisotropy constant $K_i$ being $K_{FM}$ and $K_{AFM}$, respectively. The local uniaxial anisotropy axes are set to be the in-plane direction (i.e. $\vec{e}_i = [1, 0, 0]$) for all spins to impart a uniaxial anisotropy to the simulated systems. A shape anisotropy with a negative anisotropy constant $K_i = K_{sh}$ and anisotropy axes along the out-of-plane direction (i.e. $\vec{e}_i = [0, 0, 1]$) is applied on the FM spins to simulate the effect of the demagnetization field of the FM layer. The last term describes the Zeeman energy obtained in an external field $\vec{H}$ applied along the easy-axis direction. The dimensionless units are adopted in the calculation. The exchange coupling constants $J_{FM}$ between the FM spins is defined as the unit of energy (i.e. $J_{FM} =1$). Then other parameters are chosen accordingly. $J_{AFM} = J_{INT} = -0.5 J_{FM}$ is used to ensure $T_N$ is less than $T_C$ which is preferable in the EB induced by FC. The atomic anisotropy energies $K_{FM} = 0.01 J_{FM}$, $K_{sh} = -0.05 J_{FM}$ and $K_{AFM} = 0.1 J_{FM}$ are used, since $K_{Fe}$ is on the order of 0.4 – 2.6 μ eV [38] and $K_{CoO}$ is on the order of 0.5 – 1 meV [39], where $J_{CoO}$ and $J_{Fe}$ are on the order of 6 – 10 meV [40] and 13 – 14 meV [41], respectively. The temperature $T$ was measured in units of $J_{FM}/k_B$ and the magnetic field strength $H$ in units of $J_{FM}/g\mu_B$, where $\mu_B$ is the Bohr magneton and $g$ the Landé factor [11]. All the FC $M$-$T$ curves and hysteresis loops are calculated using the standard Metropolis Monte Carlo algorithm with single spin updates. A periodic boundary condition is applied in the in-plane directions. Other details of the calculations have been described in our previous work of FM/AFM core-shell nanoparticles [42]. At each temperature and field step, the total number of Monte-Carlo steps (MCSs) used was 10000, in which 5000 initial MCSs for the thermalization were followed by 5000 MCSs for thermal averaging. The quantities show no significant dependence on further increasing MCSs in our calculations.

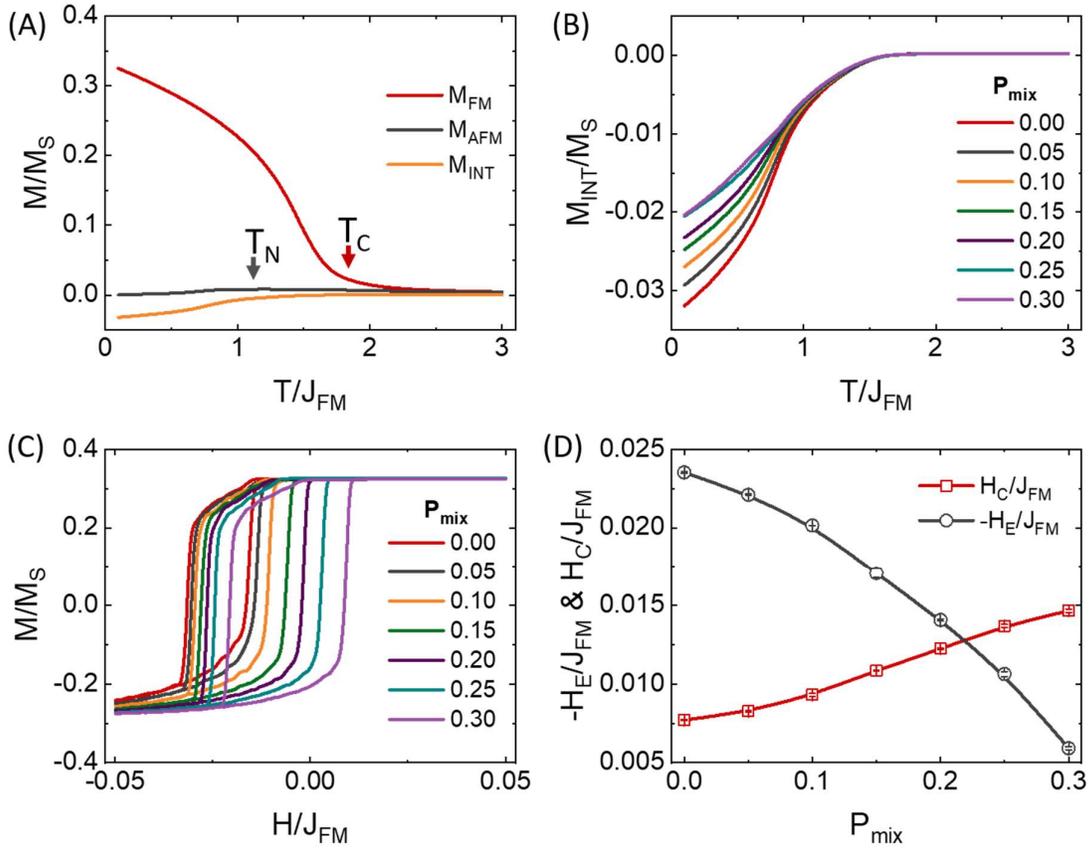

FIG. 7. (A) The FC $M$-$T$ curves of different parts of the bilayer film with $P_{mix} = 0$, (B) the FC $M$-$T$ curves of the interfacial spins, (C) hysteresis loops after FC, and (D) the extracted $H_E$

and $H_C$ obtained in bilayers with $0 \leqslant P_{mix} \leqslant 0.3$. All the data in (D) are averaged with three independent calculations with error bars coming from the calculated standard deviations.

The whole system is field cooled from $T = 3J_{FM}$ to $T = 0.1J_{FM}$ in steps of $0.1J_{FM}$ with the magnetization of each part recorded separately. Fig. 7 (A) shows the normalized *M-T* curves of the FM, AFM, and interface spins, where $M_S$ is the saturation magnetization of this whole system. As expected, a much smaller $T_N$ of the AFM compared to $T_C$ of FM is obtained. The AFM layer shows nearly a zero net magnetization after FC while the interface layer shows a net magnetization that equals the magnetization of one AFM sublattice, indicating the formation of an uncompensated interface. From Fig. 7 (B), the interfacial net magnetization $M_{INT}$ shows a steady decrease with increasing interface roughness, indicating the destruction of the uncompensated state of the interface spins by the atomic roughness. After the FC, a hysteresis loop is measured with a field range from $-0.05J_{FM}$ to $0.05J_{FM}$ with a step of $0.001J_{FM}$. Fig. 7 (C) shows the hysteresis loops with different $P_{mix}$. As the $P_{mix}$ increases, the hysteresis loops shift towards the positive field direction, which gives a decreasing $h_E$ as shown in Fig. 7 (D). The broadening of the hysteresis gives an increase of $H_C$ shown in Fig. 7 (D). These results, especially in the region $P_{mix} \geqslant 0.2$ are qualitatively in agreement with our experimental results, indicating the interface roughness plays a crucial role in the observed decrease of $H_E$ and increase of $H_C$ in the Fe/CoO (111) film with increasing $t_{CoO}$.

To find out the microscopic origin of the reduction of $H_E$ and the enhancement of $H_C$ with increase of interface roughness, the hysteresis loops of interfacial spins were studied. As shown in Fig. 8 (A), the interfacial spins present inverted hysteresis loops due to the negative interfacial exchange coupling used in our calculations. The center of all the hysteresis loops shows a vertical shift to the negative magnetization direction, indicating the existence of the uncompensated frozen interfacial spins that cannot be rotated during the FM magnetization reversals. The difference of $M_{INT}$ at the field $H_E$, i.e.

the vertical width of the hysteresis loop, contributed by the rotatable interfacial spins that can be rotated with the FM spins during the magnetization reversals, shows a steady increase with increasing $P_{mix}$. As shown in Fig. 8 (B), the magnetization contributed by the frozen spins and rotatable spins, $M_{Frozen}$ and $M_{Rotate}$ show different dependence on $P_{mix}$. Decreasing $M_{Frozen}$ and increasing $M_{Rotate}$ are observed with increasing $P_{mix}$, showing a strong correlation with the $P_{mix}$ dependence of $H_E$ and $H_C$, respectively.

For $H_E$, the contribution from the frozen interfacial spins can be roughly estimated by $H'_E = -J_{INT} M_{Frozen}/M_S$, which is calculated to be $H'_E/J_{FM}$ = -0.039 at $P_{mix}$ = 0, comparable with $H_E/J_{FM}$ = -0.024 obtained from the hysteresis loop. This indicates that the decrease in concentration of frozen spins at the uncompensated interface with increasing roughness is the reason for the decrease of the observed $H_E$.

As for $H_C$, the underlying mechanism is more complicated. The rotatable spins at the interface created by the roughness should have an effect on the observed enhancement of $H_C$ because of a larger magnetic anisotropy compared to the spins in the FM layer. The enhancement of $H_C$ from the rotatable spins can be roughly estimated from $\delta H'_C = \delta M_{Rotate} K_{AFM}/M_S$, which is calculated to be $\delta H'_C/J_{FM}$ = 0.003 with $\delta M_{Rotate}$ taken from the difference of $M_{Rotate}$ at $P_{mix}$ = 0.3 and $P_{mix}$ = 0, and which is smaller than the $\delta H_C$ = 0.0075 between the hysteresis loops with $P_{mix}$ = 0.3 and $P_{mix}$ = 0. Thus, either the roughness-induced magnetic frustration or spin compensation at the interface should also be considered in determining the enhancement of $H_C$ via pinning of the propagating domain walls or via contributions to the uniaxial anisotropy of the FM layer [31,32].

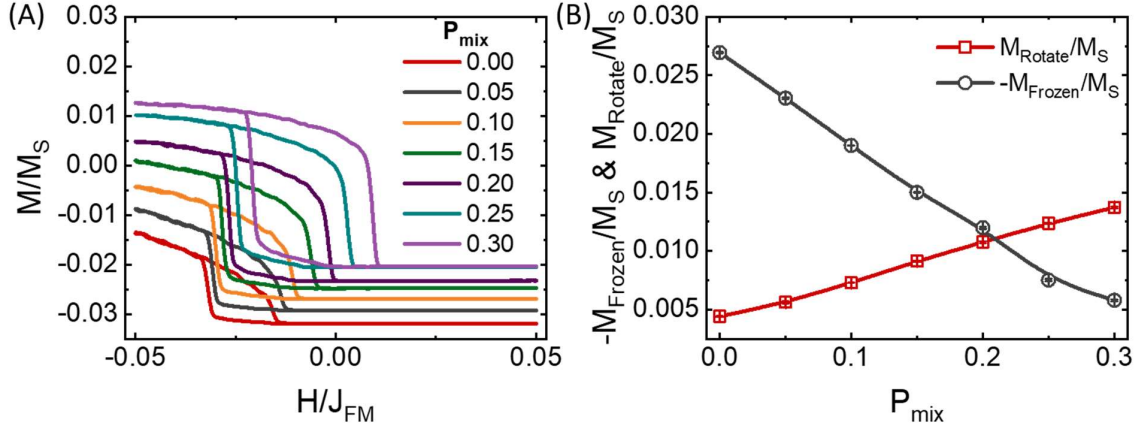

FIG. 8. (A) The hysteresis loops of the interface spins of systems with different $P_{mix}$. (B)The dependence of the magnetization from the rotatable interfacial spins and the frozen interfacial spins on $P_{mix}$.

## V.    SUMMARY AND CONCLUSIONS

In conclusion, designed interface roughness in an uncompensated Fe/CoO (111) bilayers has been induced by varying the thickness of CoO thin films grown on $Si/SiO_2$ substrates. An apparent increase of $H_C$ and a decrease of $H_E$ was observed with increasing CoO thickness, whereupon the CoO surface roughness also increased. Atomic Monte-Carlo simulations of the uncompensated FM/AFM bilayer system with varying interface roughness were also carried out. Overall, it was proven that it is the roughness at the Fe/CoO(111) interface rather than the thickness of the CoO AFM layer that plays a crucial role in determining the magnitude of the EB. On the other hand, $H_C$ was shown to be more of an interface-related property than $H_E$. This was confirmed by the determination that $T'_B \ll T_B$, $\xi'_0 \ll \xi_0$, and $\beta_s \sim 1$ in the power law $H_C \sim t^{\beta s}$. Our results represent an important step for understanding other EB systems and their uses for various device applications.

## ACKNOWLEDGMENTS

This work is supported by the National Key Research and Development Program of China (No. 2017YFA0206303, 2016YFB0700901 and 2017YFA0401502) and National Natural Science

Foundation of China (Grant Nos. 51731001 ,51371009, 11504348, 11675006), the Ph.D. Programs Foundation of Ministry of Education of China (No. 20130001110002). It was also funded by the Leverhulme Trust grant RPG-2015-017, and EPSRC grants EP/N004272/1, EP/M000524/1, and EP/L011700/1, EU grant H2020-MSCA-IF-2016 745886 MuStMAM, and SERB-DST, Government of India, Award no. SB/OS/PDF-162/2016-17.**References**